\documentclass[twocolumn]{aastex61}

\usepackage{color}

\received{}
\revised{}
\accepted{}
\submitjournal{ApJL}

\shorttitle{On the masses of clumps in distant galaxies}
\shortauthors{Dessauges-Zavadsky et al.}

\begin{document}

\title{On the stellar masses of giant clumps in distant star-forming galaxies}

\correspondingauthor{Miroslava Dessauges-Zavadsky}
\email{miroslava.dessauges@unige.ch}

\author{Miroslava Dessauges-Zavadsky}
\affil{Observatoire de Gen\`eve, Universit\'e de Gen\`eve, 51 Ch. des Maillettes, 1290 Versoix, Switzerland}

\author{Daniel Schaerer}
\affiliation{Observatoire de Gen\`eve, Universit\'e de Gen\`eve, 51 Ch. des Maillettes, 1290 Versoix, Switzerland}
\affiliation{CNRS, IRAP, 14 Avenue E. Belin, 31400 Toulouse, France}

\author{Antonio Cava}
\affiliation{Observatoire de Gen\`eve, Universit\'e de Gen\`eve, 51 Ch. des Maillettes, 1290 Versoix, Switzerland}

\author{Lucio Mayer}
\affiliation{Center for Theoretical Astrophysics and Cosmology, Institute for Computational Science, University of Zurich, Winterthurerstrasse 190, 8057 Z\"urich, Switzerland}
\affiliation{Physik-Institut, University of Zurich, Winterthurerstrasse 190, 8057 Z\"urich, Switzerland}

\author{Valentina Tamburello}
\affiliation{Center for Theoretical Astrophysics and Cosmology, Institute for Computational Science, University of Zurich, Winterthurerstrasse 190, 8057 Z\"urich, Switzerland}
\affiliation{Physik-Institut, University of Zurich, Winterthurerstrasse 190, 8057 Z\"urich, Switzerland}



\begin{abstract}

We analyse stellar masses of clumps drawn from a compilation of star-forming galaxies at $1.1<z<3.6$. Comparing clumps selected in different ways, and in lensed or blank field galaxies, we examine the effects of spatial resolution and sensitivity on the inferred stellar masses. Large differences are found, with median stellar masses ranging from $\sim 10^9~M_{\sun}$ for clumps in the often-referenced field galaxies to $\sim 10^7~M_{\sun}$ for fainter clumps selected in deep-field or lensed galaxies. We argue that the clump masses, observed in non-lensed galaxies with a limited spatial resolution of $\sim 1$~kpc, are artificially increased due to the clustering of clumps
of smaller mass. Furthermore, we show that the sensitivity threshold used for the clump selection affects the inferred masses even more strongly than resolution, biasing clumps at the low mass end. 
Both improved spatial resolution and sensitivity appear to shift the clump stellar mass distribution to lower masses, qualitatively in agreement with clump masses found in recent high-resolution simulations of disk fragmentation. We discuss the nature of the most massive clumps, and we conclude that it is currently not possible to properly establish a meaningful clump stellar mass distribution from observations and to infer the existence and value of a characteristic clump mass scale.

\end{abstract}




%

\section{Introduction}
\label{sect:introduction}

Deep {\it Hubble Space Telescope} ({\it HST}) observations and pioneering 
morphological analysis of distant star-forming galaxies have revealed that 
galaxies at the peak of the cosmic star formation activity do not follow the 
{\it Hubble} classification, but are mostly irregular and clumpy 
\citep{elmegreen05,elmegreen07,elmegreen09}. \citet{guo15} and \citet{shibuya16} 
have evaluated that at $z\gtrsim 2$ roughly 60\% of galaxies are clumpy and that 
this fraction evolves over $z\simeq 0-8$.

While the observed stellar clumps have initially been associated with 
interactions/mergers, another clump origin had to be invoked with kinematic 
studies showing a substantial proportion of $z\sim 1-2.5$ galaxies dominated by 
ordered disk rotation \citep{forster09,wisnioski15,rodrigues16}. These 
high-redshift disks, however, are very different from their local counterparts, 
being highly turbulent, thick, gas-rich, and strongly star-forming disks. They 
are subject to violent instabilities \citep{dekel09b} caused by intense inflows 
of cold gas \citep{keres05,ocvirk08,dekel09a}. Giant kpc-scale clumps with 
masses as high as $\gtrsim 10^8-10^{9.5}~M_{\sun}$ may then form during the disk 
fragmentation phase resulting from disk instabilities, as found in idealized 
simulations of isolated galaxies and cosmological simulations 
\citep{agertz09,bournaud10,bournaud14,ceverino10,ceverino12,genel12,mandelker14}. 
The produced giant clumps resemble the kpc-sized clumps observed in $z\sim 2$ 
galaxies with similar stellar masses
\citep{forster11,guo12}, and provide support to clump formation via disk 
fragmentation.

Recently, \citet[][hereafter T15]{tamburello15} and \citet{behrendt16} performed 
numerical simulations of isolated galaxies at a significantly better spatial 
resolution, 
necessary to capture fragmentation correctly \citep{mayer08}. They both find 
that the formation of giant clumps via disk fragmentation with masses 
$>10^8~M_{\sun}$ is {\em not} a common occurrence. They get the same
characteristic clump mass set by fragmentation, as low as a few times 
$10^7~M_{\sun}$, despite significant differences in their respective simulation 
techniques, 
and star formation and feedback recipes only included in T15. This fragmentation 
mass is well matched with the modified Toomre mass proposed by T15 that takes 
into account nonlinear aspects of disk fragmentation \citep{boley10}. A few 
clumps can grow to larger masses ($\sim 10^{9-9.5}~M_{\sun}$) by clump-clump 
mergers and gas accretion, but they populate only the tail of the mass 
distribution and emerge after several disk orbits. The conventional Toomre mass, 
resulting from simple linear perturbation theory \citep{toomre64}, is almost an 
order of magnitude larger, and hence appears only fortuitously comparable to the 
clump high mass tail.
A similar mass spectrum ranging from $\sim 10^{6.5}~M_{\sun}$ to 
$10^{9.5}~M_{\sun}$ is obtained for the `in situ' clumps formed in the recent 
high-resolution cosmological simulations by \citet{mandelker17}, as well as in 
the FIRE cosmological simulations by \citet{oklopcic17}.

Now that disk fragmentation simulations of different groups 
find significantly lower masses for the high-redshift clumps with respect to 
previous claims,
it is timely to revisit the observational constraints on clump masses. In this 
Letter, we compile a sample of clumps in star-forming galaxies at $1.1<z<3.6$ 
with stellar mass measurements. 
We show that a very broad range of clump masses has been derived, and find 
evidence that the derived masses and mass distributions suffer from 
limitations in both spatial resolution and sensitivity. 
We discuss what we may infer on the true stellar mass spectrum of high-redshift 
clumps. Our simple qualitative analysis presented here highlights important 
biases affecting the intrinsic clump stellar mass estimates, which we have 
started quantitatively evaluating in our first companion paper on H$\alpha$ mock 
simulations \citep[][hereafter T16]{tamburello16} and in a detailed 
observational clump analysis within a multiple-imaged galaxy (Cava et~al.\ in 
preparation).

%

\section{Clump sample}
\label{sect:samples}

We have compiled a sample of clumps from the literature within clumpy 
star-forming galaxies at $z>1$, where clumps have been identified in broad-band 
{\it HST} imaging, predominantly tracing stellar emission. Our sample comprises 
a total of 241 stellar clumps hosted in 40 galaxies from \citet{forster11}, 
\citet{guo12}, \citet{adamo13}, \citet{elmegreen13}, and \citet{wuyts14}. 
These five clump datasets are described in Table~\ref{tab:statistics}. For the 
bulk of the sample (213 out of 241 clumps), 
we have been able to recompute the clump stellar masses in a homogeneous way, 
using the original multi-band {\it HST} photometry and the updated version of 
the {\it Hyperz} photometric redshift and SED fitting code \citep{schaerer10}. 
For the remaining 28 clumps from \citet{forster11}, as observations in only one 
{\it HST}/NICMOS filter F160W are available, we have not re-analysed their 
published stellar masses, instead we rely on these estimates obtained from an 
assumed mass-to-light ratio.

The photometry of the stellar clumps from \citet{guo12} and \citet{elmegreen13} 
is based on the {\it Hubble} Ultra Deep Field observations 
\citep[HUDF,][]{beckwith06}, performed with {\it HST}/ACS in the filters F435W, 
F606W, F775W, and F850LP. \citet{guo12} also used the {\it HST}/WFC3 
observations in the filters F105W, F125W, and F160W, which were not available at 
the time of the \citet{elmegreen13} work. \citet{adamo13} have used the Cluster 
Lensing And Supernova survey with {\it Hubble} \citep[CLASH,][]{postman12} to 
analyse clumps in the filters F390W, F475W, F555W, F606W, F775W, F814W, and 
F850LP from {\it HST}/ACS, and the filters F105W, F110W, F125W, and F160W from 
{\it HST}/WFC3. And, \citet{wuyts14} had at disposal observations in the 
{\it HST}/WFC3 filters F390W, F606W, F814W, F098M, F125W, and F160W. 
For the typical redshift $z\sim 2$ of the studied clumpy host galaxies, the 
longest wavelength observations available at 1.6~$\mu$m for all cover the 
rest-frame optical emission of the stellar clumps.


For the SED modelling, we have adopted the \citet{bruzual03} stellar tracks at 
solar metallicity and the \citet{chabrier03} initial mass function. We have 
allowed for variable star formation histories, parametrised by exponentially 
declining models with timescales varying from 10~Myr to infinity\footnote{More 
precisely, we have used the following timescales $\tau=0.01$, 0.03, 0.05, 0.07, 
0.1, 0.3, 0.5, 0.7, 1., 3., $\infty$~Gyr.}, corresponding to a constant star 
formation rate. Nebular emission has been neglected, as in \citet{guo12} and 
\citet{elmegreen13}. With respect to these works, we find very small or no 
systematic differences with our inferred clump stellar masses\footnote{The mean 
of the logarithmic differences in stellar mass is 
$\Delta(\log(M^{\rm clump}_*/M_{\sun})) = -0.16\pm 0.15$ for clumps from 
\citet{guo12}, and $\Delta(\log(M^{\rm clump}_*/M_{\sun})) = +0.045\pm 0.45$ for 
clumps from \citet{elmegreen13}.}. We have also tested the impact on clump 
stellar masses when including or not the near-infrared {\it HST}/WFC3 photometry 
in the dataset of \citet{guo12}. We find higher stellar masses by +0.20~dex, on 
average, when the {\it HST}/WFC3 filters are omitted, as done in 
\citet{elmegreen13}. This could thus lead to a small systematic shift by a 
factor of $\sim 1.5$ between the clump masses of \citet{guo12} and 
\citet{elmegreen13}.
Compared to the clump stellar masses reported by \citet{adamo13}, our masses are 
higher by +0.56~dex, on average. This difference 
vanishes\footnote{$\Delta(\log(M^{\rm clump}_*/M_{\sun})) = +0.004\pm 0.46$.} if 
we include nebular emission and allow for ages younger than 10~Myr in the SED 
fits, as adopted by \citet{adamo13}. For the \citet{wuyts14} dataset, we find a 
small difference of $-0.12$~dex, on average, in clump stellar masses when 
neglecting nebular emission, as in their work. For a uniform and conservative 
comparison between all the clump datasets, we retain the clump stellar masses 
obtained from SED fits {\it without} nebular emission.


The wavelength coverage of the above four clump datasets is nearly identical, 
which thus enables a meaningful and nearly homogeneous comparison between these 
datasets. As a measure of the depth of the selected clumps, we examine the clump 
magnitude distributions and we list in Table~\ref{tab:statistics}, for the four 
clump datasets, the magnitudes {\it i}$_{16}$ and {\it z}$_{16}$ corresponding to 
the 16{\it th} percentile of the magnitude distribution of clumps
in the F775W (for HUDF) or F814W {\it i}-band and in the F850LP {\it z}-band, 
respectively. The latter corresponds to the clump selection band of 
\citet{guo12}, and the former is the second-deepest band in HUDF and CLASH. We 
also indicate the $3\,\sigma$ sensitivity limits in the {\it i}- and 
{\it z}-bands measured in 0.35\arcsec\ diameter apertures, as reported by 
\citet{beckwith06} for HUDF and \citet{postman12} for CLASH. 
On this basis, we divide the clumpy host galaxy compilation into three main 
sub-samples, lensed galaxies, field galaxies with a deep clump selection, and 
field galaxies with a shallow clump selection, denoted hereafter by L, FD, and 
FS, respectively (see Table~\ref{tab:statistics}). 

The 40 host galaxies have redshifts ranging from $z=1.1$ to 3.6, with the bulk 
found at $1.3<z<2.6$. Their stellar masses are uniformly distributed between 
$M_*^{\rm host} \sim 10^8-10^{11}~M_{\sun}$ (see Figure~\ref{fig:clump-host}), 
with the L and FD sub-samples containing the low-mass host galaxies 
($M_*^{\rm host} \lesssim 10^{10}~M_{\sun}$) and the FS sub-sample the high-mass 
hosts ($M_*^{\rm host} \gtrsim 10^{9.8}~M_{\sun}$). Most of the hosts are on the 
main sequence at their corresponding redshift.


For comparison we also consider local star clusters found in nearby galaxies 
\citep{adamo13} and two starburst galaxies \citep{bastian06,larsen02}. 


%

\begin{deluxetable*}{lcc|c|cc}
\tablecolumns{6}
\tablecaption{Properties of existing stellar clump datasets in high-redshift galaxies \label{tab:statistics}}
\tablewidth{0pt}
\tablehead{
\colhead{References} & \colhead{Adamo+13} & \colhead{Wuyts+14} & \colhead{Elmegreen+13} & \colhead{Guo+12} & \colhead{F\"orster Schreiber+11} \\
 & \multicolumn{2}{c}{L sub-sample\tablenotemark{a}} & \colhead{FD sub-sample\tablenotemark{b}} & \multicolumn{2}{c}{FS sub-sample\tablenotemark{c}} 
}
\startdata
Number of clumps                                           & 31 & 7 & 135 & 40 & 28  \\ 
Number of host galaxies                                    & 1  & 1 & 22  & 10 & 6  \\
Redshift                                                   & 1.5 & 1.7 & $1.1-3.6$ & $1.6-2.0$ & $2.2-2.5$ \\
{\it i}$_{16}$\tablenotemark{d}                            & 29.7\tablenotemark{$\dag$} & 29.1\tablenotemark{$\ddag$} & 29.7  & 27.6  & -- \\
{\it i}-band $3\sigma$-0.35\arcsec\ limit\tablenotemark{e} & 30.5\tablenotemark{$\dag$} & 30.9\tablenotemark{$\ddag$} & 30.25 & 30.25 & -- \\
{\it z}$_{16}$\tablenotemark{d}                            & 29.7\tablenotemark{$\dag$} & --                          & 29.7  & 27.3  & -- \\
{\it z}-band $3\sigma$-0.35\arcsec\ limit\tablenotemark{e} & 29.5\tablenotemark{$\dag$} & --                          & 29.55 & 29.55 & -- \\
Median $\log(M_*^{\rm clump}/M_{\sun})$ & \multicolumn{2}{c|}{6.98} & 7.23 & \multicolumn{2}{c}{8.89} \\ 
\enddata
\tablenotetext{a}{Clumps identified in lensed galaxies.} 
\tablenotetext{b}{Clumps identified in field galaxies with a deep clump selection.}
\tablenotetext{c}{Clumps identified in field galaxies with a shallow clump selection.}
\tablenotetext{d}{Magnitudes corresponding to the 16{\it th} percentile of the magnitude distribution of clumps in the F775W (for HUDF) or F814W {\it i}-band and in the F850LP {\it z}-band, respectively.}
\tablenotetext{e}{$3\,\sigma$ sensitivity limits in the {\it i}-, respectively, {\it z}-band measured in 0.35\arcsec\ diameter apertures (from \citet{beckwith06} for HUDF and \citet{postman12} for CLASH).}
\tablenotetext{$\dag$}{Corrected for lensing, assuming a magnification factor $\mu=8$ \citep{adamo13}.}
\tablenotetext{$\ddag$}{Corrected for lensing, assuming a magnification factor $\mu=25$ \citep{sharon12}.}
\end{deluxetable*}
%

\begin{figure}
\centering
\includegraphics[width=8.3cm,clip]{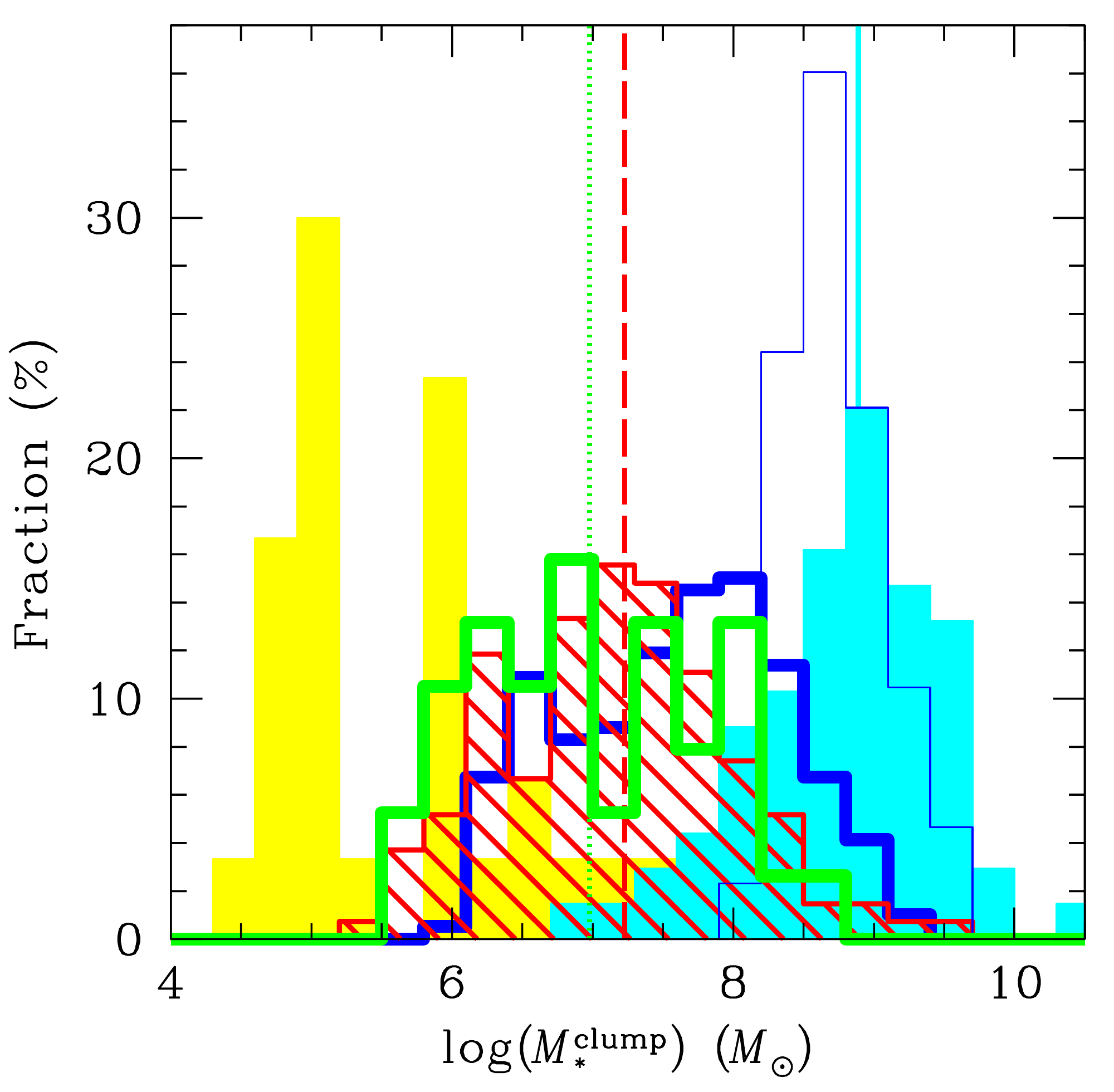}
\caption{Normalized stellar mass distributions of local star clusters (filled yellow histogram), and three sub-samples of high-redshift clumps: clumps in lensed galaxies (L sub-sample, open green histogram), in field galaxies with a deep clump selection (FD sub-sample, hatched red histogram), and in field galaxies with a shallow clump selection (FS sub-sample, filled cyan histogram). The medians of the high-redshift clump sub-samples are shown using dotted green, dashed red, and solid cyan vertical lines, respectively. For comparison, clump mass distributions as predicted by different disk fragmentation simulations (open blue thick and thin histograms from T15 and \citet{ceverino12}, respectively) are also shown in each panel.} 
\label{fig:clump-M*}
\end{figure}
%

\section{Can we infer accurate clump masses at high redshift?}
\label{sect:results}

As shown in Figure~\ref{fig:clump-M*}, the distribution of stellar masses of 
clumps identified in high-redshift galaxies is very broad and ranges from 
$M_*^{\rm clump}\sim 10^{5.5}~M_{\sun}$ to $10^{10.5}~M_{\sun}$. Large 
differences are observed among the three sub-samples of high-redshift galaxies 
considered here. Whereas the ``typical'' mass of clump masses in the field 
galaxies studied by \citet{forster11} and \citet{guo12} (FS sub-sample)~--~used 
until now as the benchmark of high-redshift clump properties~--~is very high 
(median $\log (M_*^{\rm clump,FS}/M_{\sun}) = 8.89$), the \citet{elmegreen13} 
field galaxies (FD sub-sample) have a median clump mass much lower 
($\log (M_*^{\rm clump,FD}/M_{\sun}) = 7.23$), and clumps in lensed galaxies 
(L sub-sample) show even somewhat lower masses (median 
$\log (M_*^{\rm clump,FS}/M_{\sun}) = 6.98$; see Table~\ref{tab:statistics}). 
In comparison to the star clusters identified in local galaxies, the inferred 
clump masses in high-redshift galaxies are, on average, significantly higher 
than those in local galaxies also shown in Figure~\ref{fig:clump-M*}, with the 
exception of some star clusters in the most intensively star-forming nearby 
galaxies.

\begin{figure}
\centering
\includegraphics[width=8.3cm,clip]{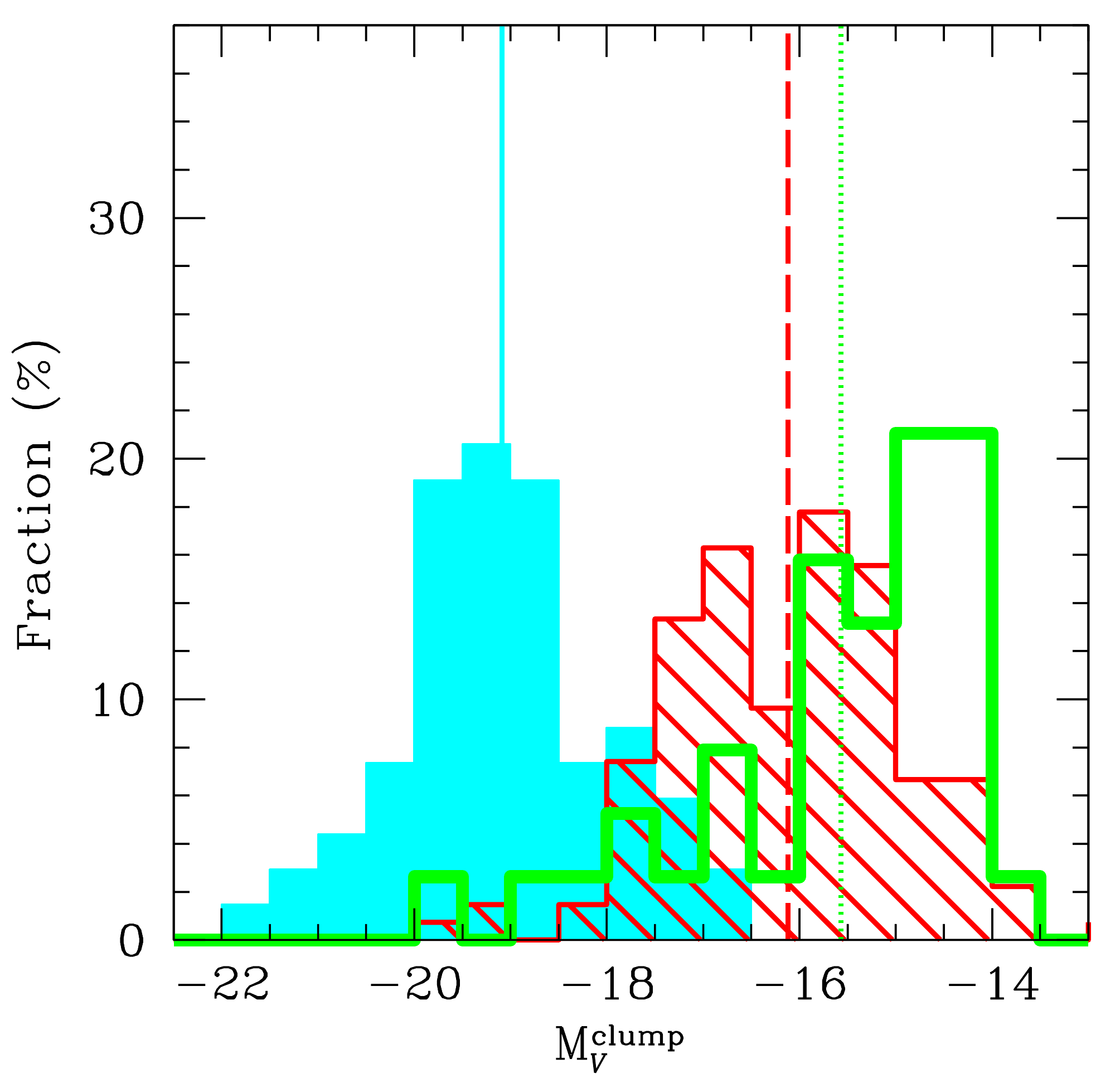}
\caption{Absolute rest-frame {\it V}-band magnitude distributions of high-redshift clumps in the L (open green histogram), FD (hatched red histogram), and FS (filled cyan histogram) sub-samples. The respective means are shown using dotted green, dashed red, and solid cyan vertical lines.} 
\label{fig:clump-Mag}
\end{figure}

The absolute rest-frame {\it V}-band magnitude distributions of the three clump 
sub-samples are compared in Figure~\ref{fig:clump-Mag}. Clearly, the FS 
sub-sample has significantly brighter clumps than the FD sub-sample, although 
both are drawn from field galaxies over a similar redshift range. The clumps in 
the lensed galaxies are slightly fainter, on average, than those in the FD 
sub-sample. The differences in absolute magnitude and in stellar mass 
(Figure~\ref{fig:clump-Mag} versus Figure~\ref{fig:clump-M*}) are comparable, as 
expected, since the optical light traces stellar mass if the mass-to-light ratio 
of clumps does not vary much.

What explains the large differences found between the three sub-samples of 
high-redshift clumps? We primarily envisage {\em spatial resolution} and 
{\em sensitivity} as the main sources for these differences. 

All high-redshift clumps rely on {\it HST} imaging with the same spatial 
resolution of $\sim 0.15\arcsec$ FWHM, which corresponds to physical sizes of 
$1.2-1.3$~kpc at $z\sim 1.3-2.6$ in field/non-lensed galaxies.
Obviously, limited spatial resolution can affect the measure of clump stellar 
masses, if the true physical sizes of clumps are smaller than the resolution, 
since then several clumps may be blended within the photometric aperture. This 
effect will artificially ``boost'' the flux and increase the inferred stellar 
mass of clumps. The amount of this artificial boost will depend on the clump 
true sizes, their distribution and clustering. 
With the help of strong gravitational lensing,
sub-kpc sizes down to $\sim 100$~pc (representing an improvement by a factor of 
10) are reached 
in the two lensed galaxies of the L sub-sample.
The finding of considerably lower clump stellar masses 
(Figure~\ref{fig:clump-M*} and Table~\ref{tab:statistics}) compared to the 
clump masses in the field galaxy sub-sample(s) (FS and somewhat FD) supports 
that indeed spatial resolution, and the induced blending, affects the derived 
clump masses and artificially boosts them towards high masses. 

%

%

First quantitative hints of this low-resolution ``boosting'' on clump stellar 
masses have been obtained from recent simulations, which, however, focus on gas 
clumps. \citet{behrendt16}, in their simulations of a massive gas disk with one 
of the highest resolutions to date, find very small ($\sim 35$~pc in radii) and 
low gas mass fragments produced 
with disk fragmentation
that, when mimicking observations on kpc-scales ($\rm FWHM=1.6~kpc$), appear to 
be distributed in loosely bound clusters 
with $10-100$ times larger masses.
We report similar results in T16 using our H$\alpha$ mock observations of 
simulations from T15, and we infer a $\sim 1$~kpc resolution ``boosting'' on 
100~pc-scale clump masses of less than a factor of 5.
Apart from that, \citet{fisher17},
using low-redshift H$\alpha$ galaxy observations, have analysed how severely 
clump clustering increases sizes and star formation rates in limited 
$\sim 1$~kpc resolution maps.

Interestingly, large stellar mass differences are also observed between clumps 
in the field galaxy FS and FD sub-samples (Figure~\ref{fig:clump-M*} and 
Table~\ref{tab:statistics}), while these galaxies are all affected by the same 
$\sim 1.2$~kpc resolution limitation. Another effect than spatial resolution 
must thus be at the origin of these clump mass differences.
These differences are likely due to different clump selections applied, 
resulting from different data depths, different wavebands used to identify the 
clumps, and/or more or less conservative 
detection limits set for clumps. In fact both \citet{guo12}\footnote{The clumps 
from \citet{guo12} represent 60\% of the clumps in the field galaxy FS 
sub-sample.} 
and \citet{elmegreen13} used HUDF observations, but the former selected clumps 
in the F850LP {\it z}-band, whereas the latter in the F775W {\it i}-band that is 
0.7~mag deeper (see Table~\ref{tab:statistics}).
Furthermore, the clumps extracted by \citet{guo12} are limited to F850LP 
magnitudes brighter than $\sim 27.3$, well above the depth of the HUDF 
{\it z}-band image.
In contrast, the observed magnitudes of clumps selected by \citet{elmegreen13} 
reach down to $3\,\sigma$, which can explain differences of up to $\sim 2.5$~mag 
for the faintest clumps in the FD sub-sample (see Table~\ref{tab:statistics}) 
compared to \citet{guo12}. Hence, the clump selection sensitivity threshold 
strongly affects the clump stellar masses, biasing the observed clumps at the 
low mass end.

The sensitivity effect appears to be more important than the spatial resolution 
effect on the inferred clump masses, 
since the respective stellar mass distributions of clumps in the
\citet{elmegreen13} field galaxies limited by $\sim 1.2$~kpc resolution 
(FD sub-sample) and in the lensed galaxies (L sub-sample) end up to be very 
comparable
(Figure~\ref{fig:clump-M*} and Table~\ref{tab:statistics}), whereas clumps in 
the lensed galaxies benefit from 10 times better spatial resolution {\it and} 
similarly good sensitivities.

In any case, the finding of clumps in lensed galaxies and in field galaxies from 
\citet{elmegreen13} with stellar masses between $\sim 10^{5.5}-10^9~M_{\sun}$, 
well below the often-quoted ``typical'' masses of giant clumps 
$\gtrsim 10^8-10^9~M_{\sun}$ inferred from observations with $\sim 1.2$~kpc 
resolution and shallower clump selection thresholds (FS sub-sample), suggests 
that the latter is systematically overestimated 
by $1-2$ orders of magnitude (Table~\ref{tab:statistics}), or more depending on 
whether a characteristic mass scale of fragmentation exists or not (see 
Section~\ref{sect:spectrum}). The same conclusion can be drawn, when we restrict 
the FD and FS sub-samples to host galaxies with redshifts comparable to those of 
the two lensed galaxies from the L sub-sample.
In T16 we study quantitatively the effects of $\sim 1$~kpc resolution and 
shallow sensitivity on the observed clump masses using H$\alpha$ mocks. 
We find that the inferred clump stellar masses can be easily overestimated by at 
least a factor of 10 due to the combination of both effects (and with the 
sensitivity effect dominating). 

%

\section{Discussion}
\label{sect:discussion}



%

\subsection{On the existence of the most massive clumps}
\label{sect:max}

\begin{figure}
\centering
\includegraphics[width=8.3cm,clip]{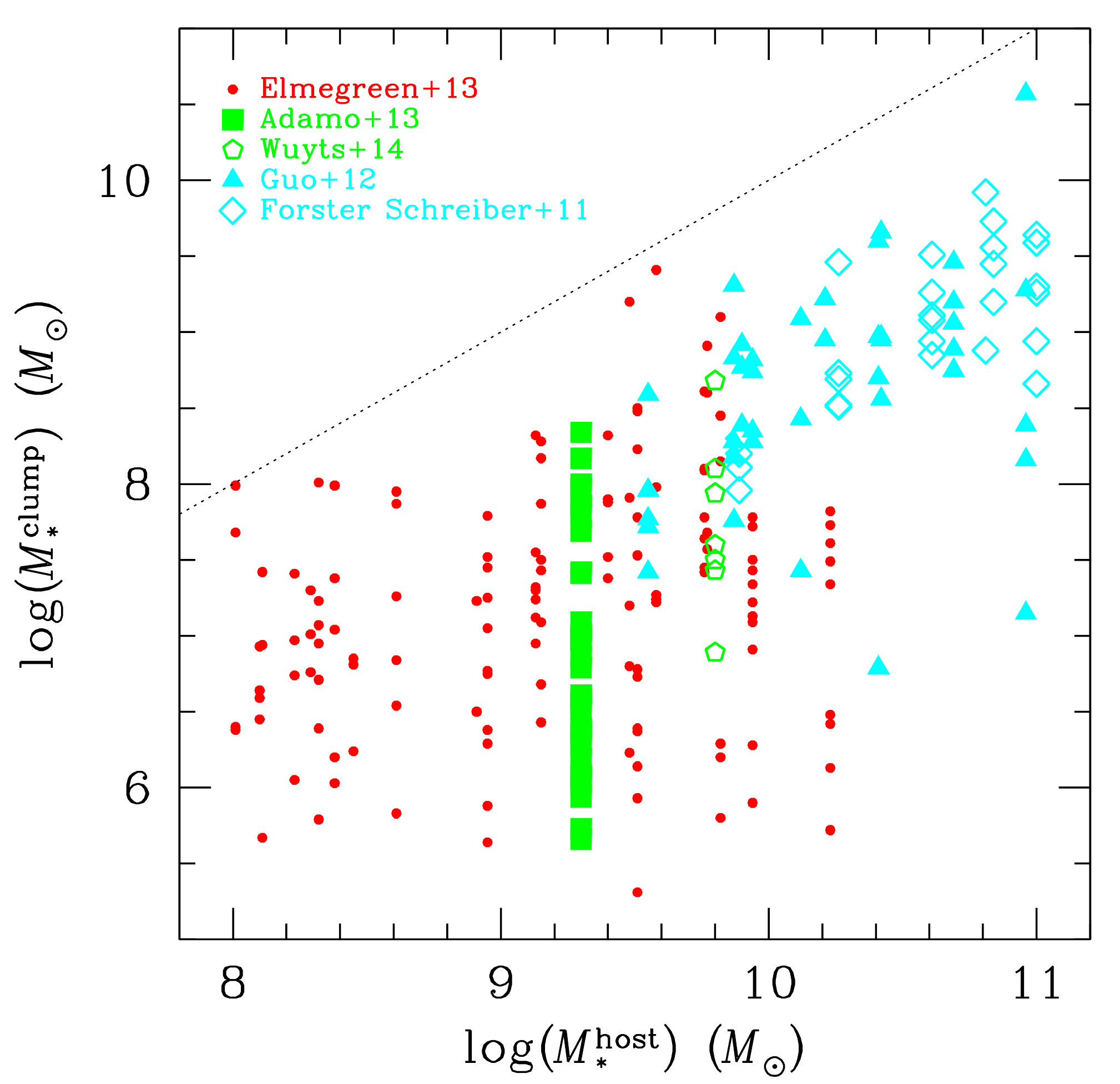}\hspace{1cm}
\caption{Stellar masses of high-redshift clumps plotted as a function of the stellar mass of their host galaxy. The symbols refer to different works, and the color-coding to the L, FD, and FS sub-samples, similarly to Figures~\ref{fig:clump-M*} and \ref{fig:clump-Mag}. The dotted line is the one-to-one relation.}
\label{fig:clump-host}
\end{figure}

Is there a maximum stellar mass for clumps, how massive, and what determines 
it? If clump stellar masses are artificially increased by the spatial 
resolution effect 
as discussed above, our current best maximum clump mass estimate should come 
from lensed galaxies, where clump stellar masses up to $\sim 10^{8.7}~M_{\sun}$ 
are observed (see Figure~\ref{fig:clump-M*}). However, the L sub-sample is quite 
small (38 clumps) and small number statistics could bias the maximum mass 
determination of clumps (especially if the true clump mass function decreases 
rapidly towards high masses). Furthermore, the maximum clump stellar mass could 
depend on the host galaxy stellar mass 
\citep[see][]{elmegreen13}. 
The fact that clumps in the FS and FD sub-samples, observed with the same 
spatial resolution in host galaxies spanning a wide range of stellar masses, 
show an increase of the upper envelope of their stellar masses with the host 
galaxy mass as illustrated in Figure~\ref{fig:clump-host}, indicates that the 
maximum clump mass indeed depends on the host mass. 

By definition, the clump mass cannot exceed the host galaxy mass, but what 
determines the maximum clump mass? The most simple expectation is that the 
maximum clump mass is set by the fragmentation mass that is directly 
proportional to the galaxy mass and the square of its gas fraction in the linear 
perturbation theory, as described by \citet{escala08}. Otherwise, according to 
the innovative simulations of T15, the combination of a typical fragmentation 
scale and additional processes yielding the clump mass growth, such as 
clump-clump mergers, leads to a fractional stellar mass contribution of the sum 
of all clumps to the total disk stellar mass in the range of $10-15$\%, with 
little variation with disk mass. 
This results from the fact that the characteristic mass scale of fragmentation 
they get is independent on disk mass. 
Massive disks thus just give rise to more clumps that in turn increase the 
likelihood of clump-clump mergers, shifting the maximum stellar mass of clumps 
to larger values. 
Both approaches allow to explain the apparent scaling of the maximum clump 
mass with the host galaxy mass.

%

%

On the other hand, we could expect the clump properties to correlate with 
redshift, such that the more massive clumps should be found in the higher 
redshift host galaxies, since both the velocity rotation over dispersion ratio 
and the molecular gas fraction, which together control the Toomre disk stability 
criterion, have been shown to increase with redshift
\citep{wisnioski15,dessauges15}. However, no such a trend is observed, when plotting the measured clump stellar masses as a function of the redshift of their hosts.


If we assume that the simulations of T15 and \citet{mandelker17} predict 
correctly the stellar masses of the order of $\sim 10^{9-9.5}~M_{\sun}$
of the most massive clumps\footnote{This maximum clump mass is obtained for 
simulated galaxies with stellar masses up to $10^{10.6}~M_{\sun}$ (T15),
comparable to the field galaxies in the FS sub-sample ($M_*^{\rm host} \gtrsim 
10^{9.8}~M_{\sun}$). Limiting the simulations of T15 to galaxies with 
$M_*^{\rm host} < 10^{10}~M_{\sun}$ (comparable to galaxies in the FD and L 
sub-samples), leads to a clump stellar mass distribution not exceeding 
$\sim 10^{8.5}~M_{\sun}$.}, we see that still a non-negligible fraction 
of the observed clumps in the FS sub-sample has stellar masses above the 
$10^{9.5}~M_{\sun}$ limit (Figures~\ref{fig:clump-M*} and \ref{fig:clump-host}). 
Explaining these extreme clump masses with the spatial resolution effect appears 
difficult as several very massive clumps would need to be closely clustered.
In our H$\alpha$ mocks (T16), maximum stellar masses up to $\sim 3\times 
10^9~M_{\sun}$ can be reached in artificially inflated $\sim 1$~kpc clumps. 
When examining these extremely massive clumps from \citet{forster11} and 
\citet{guo12} individually, we find that almost all of them coincide with the 
centers of host galaxies or are located very close by, and have among the 
reddest colors. They thus appear more suggestive of galactic bulges, or alike, 
rather than genuine clumps \citep[see also][]{elmegreen09}. But, it has also been 
proposed that they could be old clumps that have migrated in the centers of 
galaxies \citep{wuyts12}. Their extreme masses remain, nevertheless, puzzling. 
Contributions from other processes than disk fragmentation and clump-clump 
mergers that follow, such as minor mergers or accretion of cores of disrupted 
satellites \citep[Ribeiro et~al.\ 2016, in preparation;][]{mandelker17}, can be 
an alternative way to explain star complexes with extreme masses, 
eventually red colors, and central galaxy positions after migration.

%

\subsection{Is there a characteristic clump mass from observations?}
\label{sect:spectrum}


As shown in Section~\ref{sect:results}, {\it HST} imaging has revealed 
high-redshift clumps with a wide range of stellar masses,
typically spread over two orders of magnitude, or significantly larger if data 
with different sensitivities and spatial resolutions are combined 
(Figure~\ref{fig:clump-M*}). Furthermore, in each clump dataset the lower 
stellar mass end is limited by the depth and spatial resolution of the available 
observations. From this we conclude that it is currently not possible to 
properly establish a meaningful clump stellar mass distribution from 
observations, and, in particular, to infer the existence and value of a 
characteristic clump mass. The only clear indication is that both improved 
sensitivity and spatial resolution shift the clump stellar mass distribution to 
lower masses that ends up to be in agreement with the latest simulations of disk 
fragmentation. Indeed, 
T15 find a characteristic clump stellar mass of $\sim 5\times 10^7~M_{\sun}$ at 
the onset of fragmentation and predict a stellar mass distribution of clumps, 
also plotted in Figure~\ref{fig:clump-M*}, which very much resembles that of 
clumps in 
the L and FD sub-samples. The agreement may well be fortuitous for the reasons 
just discussed. 

If clumps are formed by disk fragmentation and molecular clouds down to several 
orders of magnitude lower mass scales are formed primarily by the same mechanism 
\citep[e.g.,][]{tasker09,krumholz10}, clump formation would be hierarchical and, 
hence, one would expect clumps to continuously reveal new substructure at all 
scales \citep{elmegreen11,bournaud16}, making it impossible to assess their mass 
distribution in a resolution-independent way. 
Observational evidence for a hierarchical star cluster structure in nearby 
galaxies is discussed by \citet{gouliermis15}. On the other end, if high-
redshift disks do possess a characteristic fragmentation mass scale as suggested 
by simulations of T15 and \citet{behrendt16}, the signature of such scale should 
be independent on spatial resolution and sensitivity once observations approach 
the corresponding scale. Convergence studies of simulations with increased 
resolution will help assess the latter mass scale robustly in the context of the 
fragmentation scenario.
At the same time, larger high-redshift clump samples within deep observations, 
ideally at the best-possible spatial resolution, and a systematic analysis (with 
the same clump selection criteria), including completeness corrections, are 
needed to establish the true clump stellar mass spectrum. 

\acknowledgments

This work was supported by the Swiss National Science Foundation, in the context 
of the Sinergia STARFORM network on ``Star formation in galaxies from the 
Milky Way to the distant Universe''. We are grateful to Bruce G. Elmegreen and 
Yicheng Guo for sharing with us their {\it HST} clump photometry, and we warmly 
thank Bruce G. Elmegreen for very fruitful discussions.

%



\end{document}